\documentclass[
    twocolumn, superscriptaddress, 
    amsmath,amssymb, aps]{revtex4-2}%
    
\usepackage{graphicx}
\usepackage{dcolumn}
\usepackage{bm}

\usepackage{physics}
\usepackage[hidelinks]{hyperref}
\usepackage[dvipsnames]{xcolor}
\usepackage{mathtools}
\usepackage[normalem]{ulem} 

\begin{document}

\title{Programmable linear quantum networks with a multimode fibre}

\author{Saroch Leedumrongwatthanakun}
    \affiliation{Laboratoire Kastler Brossel, ENS-PSL Research University, CNRS, Sorbonne Universit\'e, Coll\`ege de France, 24 rue Lhomond, Paris 75005, France}
\author{Luca Innocenti}%
    \affiliation{Centre for Theoretical Atomic, Molecular, and Optical Physics, School of Mathematics and Physics, Queen’s University Belfast, BT7 1NN Belfast, United Kingdom}

\author{Hugo Defienne}
    \affiliation{Laboratoire Kastler Brossel, ENS-PSL Research University, CNRS, Sorbonne Universit\'e, Coll\`ege de France, 24 rue Lhomond, Paris 75005, France}
    \affiliation{School of Physics and Astronomy, University of Glasgow, Glasgow G128QQ, United Kingdom }

\author{Thomas Juffmann}
    \affiliation{Laboratoire Kastler Brossel, ENS-PSL Research University, CNRS, Sorbonne Universit\'e, Coll\`ege de France, 24 rue Lhomond, Paris 75005, France}
    \affiliation{Faculty of Physics, University of Vienna, A-1090 Vienna, Austria}
    \affiliation{Department of Structural and Computational Biology, Max F. Perutz Laboratories, University of Vienna, A-1030 Vienna, Austria}
    
\author{Alessandro Ferraro}
    \affiliation{Centre for Theoretical Atomic, Molecular, and Optical Physics, School of Mathematics and Physics, Queen’s University Belfast, BT7 1NN Belfast, United Kingdom}%
    
\author{Mauro Paternostro}
    \affiliation{Centre for Theoretical Atomic, Molecular, and Optical Physics, School of Mathematics and Physics, Queen’s University Belfast, BT7 1NN Belfast, United Kingdom}%
    
\author{Sylvain Gigan}
 \email{sylvain.gigan@lkb.ens.fr}
    \affiliation{Laboratoire Kastler Brossel, ENS-PSL Research University, CNRS, Sorbonne Universit\'e, Coll\`ege de France, 24 rue Lhomond, Paris 75005, France}%

\begin{abstract}
Reconfigurable quantum circuits are fundamental building blocks for the implementation of scalable quantum technologies. Their implementation has been pursued in linear optics through the engineering of sophisticated interferometers~\cite{OBrien2007,Matthews2009,Carolan2015}. While such optical networks have been successful in demonstrating the control of small-scale quantum circuits, scaling up to larger dimensions poses significant challenges~\cite{Flamini2018,Harris2018}. Here, we demonstrate a potentially scalable route towards reconfigurable optical networks based on the use of a multimode fibre and advanced wavefront-shaping techniques. We program networks involving spatial and polarisation modes of the fibre and experimentally validate the accuracy and robustness of our approach using two-photon quantum states. In particular, we illustrate the reconfigurability of our platform by emulating a tunable coherent absorption experiment~\cite{Baranov2017}. By demonstrating reliable reprogrammable linear transformations, with the prospect to scale, our results highlight the potential of complex media driven by wavefront shaping for quantum information processing.
\end{abstract}
   
\maketitle
Linear optical networks are prominent candidates for practical quantum computing~\cite{OBrien2007}. The efficient implementation of quantum information processing tasks requires high dimensionality, dense network connectivity and the possibility to actively reconfigure the network. Currently, bulk and integrated linear optics are the most popular platforms to implement such networks. The design of the latter is based on a cascade of beamsplitters and phase-shifters connected by single-mode waveguides~\cite{Matthews2009,Carolan2015,Flamini2018,Harris2018}. However, the scalability of such architecture is significantly limited by the fabrication process. Alternatively, integrated multimode waveguides~\cite{Peruzzo2011,Poem2012,Feng2016,Mohanty2017} and metasurfaces~\cite{Wang2018} provided new routes towards robust implementation of larger quantum optical circuits, with the strong disadvantages of not being reprogrammable after fabrication. Coupling spatial modes with other degrees of freedom, such as time, frequency and polarisation~\cite{Xu2014a}, provides a different route towards encoding and processing information in higher dimensions~\cite{Lanyon2009}, but remains an engineering challenge in integrated optics. To date, the quest for a controllable high-dimensional optical network offering arbitrary connectivity is ongoing.

Complex media, from white paint to multimode fibres, can overcome these bottlenecks when used in combination with wavefront shaping. Many classical and quantum applications rely on this approach~\cite{Rotter2017}, ranging from spatial mode structuring~\cite{Morizur2010,Fickler2017,Wang2017} to adaptive quantum optics~\cite{Defienne2018}. As for linear circuits, programmable beamsplitters have been implemented in opaque scattering media~\cite{Huisman2014,Huisman2015,Wolterink2016} and multimode fibres~\cite{Defienne2016} through control of spatial mode mixing. In this work, we report the implementation of fully programmable linear optical networks of higher dimensions by harnessing spatial and polarisation mixing processes in a multimode fibre driven by wavefront shaping. We first demonstrate the reliability and versatility of our approach by controlling two-photon interferences between multiple ports of various networks with high accuracy. We then emulate a circuit for tunable coherent absorption, which highlights the reconfigurable nature of our platform. Our work demonstrates the viability of coherent manipulation of optically encoded information via multimode scattering from complex media and wavefront shaping, and its potential for quantum information processing.
\begin{figure*}[htp]
\centering
\includegraphics[width=\linewidth]{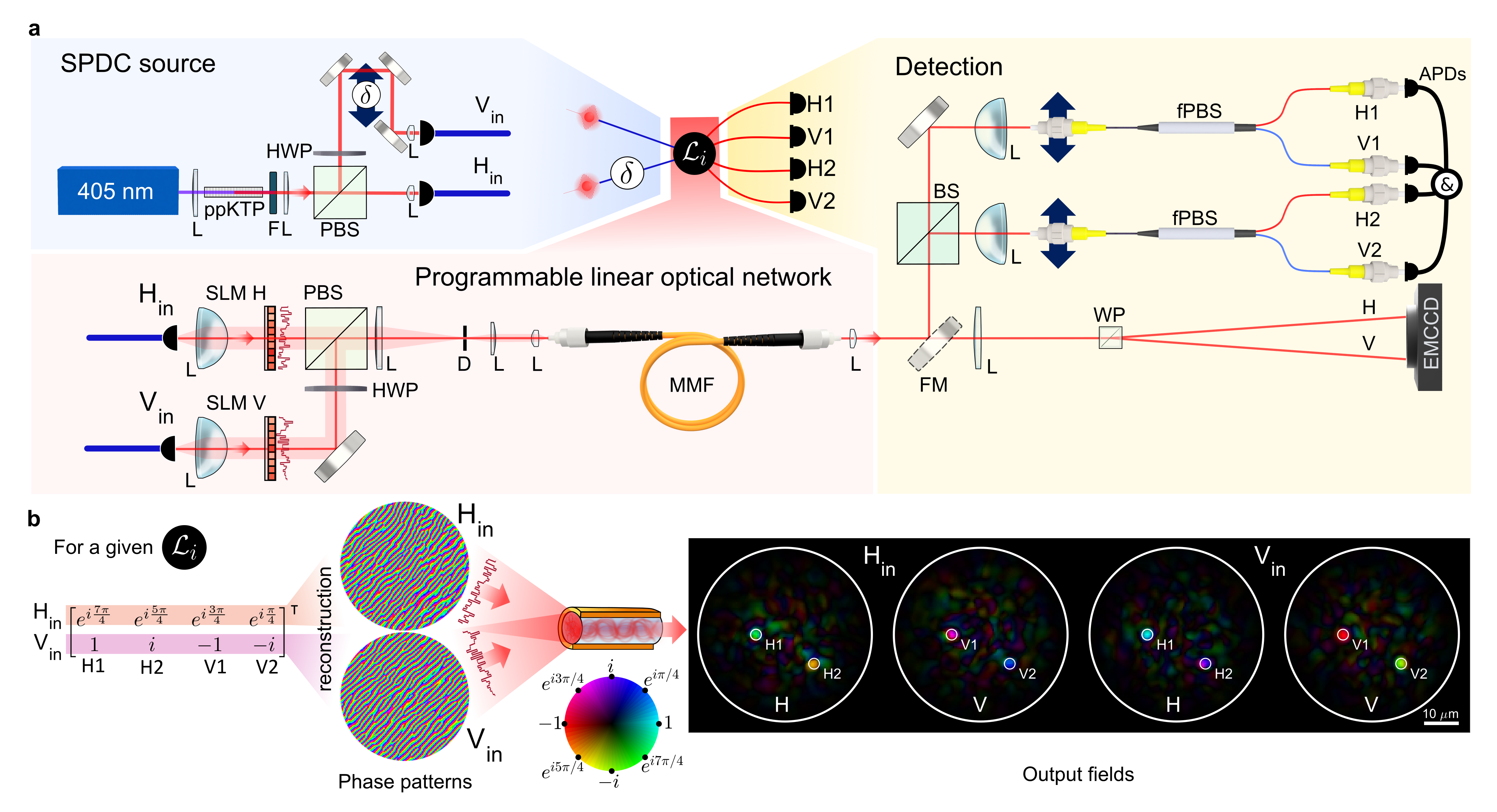}
\caption{\textbf{Multimode-fibre based programmable linear optical network.} (a) Conceptual schematics of the apparatus. Photon pairs produced by SPDC are injected into an MMF along orthogonal polarisation using SLMs. We use commercial MMF (Thorlabs, GIF50C) as a tool to achieve mode mixing. The TM is measured across spatial and polarisation modes of the MMF (see SI Section I). The wavefront corresponding to a desired linear transformation $\mathcal{L}_{i}$ is calculated and displayed on the SLMs (see Methods). Output ports of interest are selected by two single-mode fibre-based polarisation beamsplitters (fPBS) mounted on translation stages. These correspond to two spatial modes and two polarisations labelled as (H1, V1, H2, V2). Light is detected by avalanche photodiodes (APDs) connected to a coincidence electronics. The output plane of the MMF is imaged onto an electron multiplying charge-coupled device (EMCCD) camera along both polarisations (H and V). (b) An arbitrary $4 \times 2$ linear network $\mathcal{L}_{i}$ is implemented by shaping the spatial phases of each input port $H_{in}$ and $V_{in}$. For each input, the predicted output fields after propagation through the MMF are shown. We observe that light is focused into the four targeted output ports with the desired amplitudes and phases. (L: lenses; F: filter; HWP: half wave plate; PBS: polarising beamsplitter; D: iris diaphragm; FM: flip mirror; WP: Wollaston prism; BS: beamsplitter; ppKTP, periodically poled potassium titanyl phosphate crystal).}
\label{fig1}
\end{figure*}

The experiment is conceptually illustrated in Fig.~\ref{fig1}. The multimode fibre (MMF) is a graded-index fibre supporting $\sim400$ propagation modes at $\lambda=$ 810 nm. Complex spatial and polarisation mixing occurring in the fibre is the key ingredient that enables the design of a reconfigurable linear transformation $\mathcal{L}_i$. Indeed, measuring the transmission matrix (TM) of the MMF reveals its highly isotropic connectivity across spatial and polarisation modes (see Supplementary Information (SI) Section 1-2). We exploit the connectivity together with the near-unitary of the MMF to program linear optical transformations $\mathcal{L}_i$ (see Methods for details) in a four-dimensional Hilbert space defined across spatial and polarisation degrees of freedom, labelled H1, V1, H2, V2.

\begin{figure}[htp]
\centering
\includegraphics[width=\linewidth]{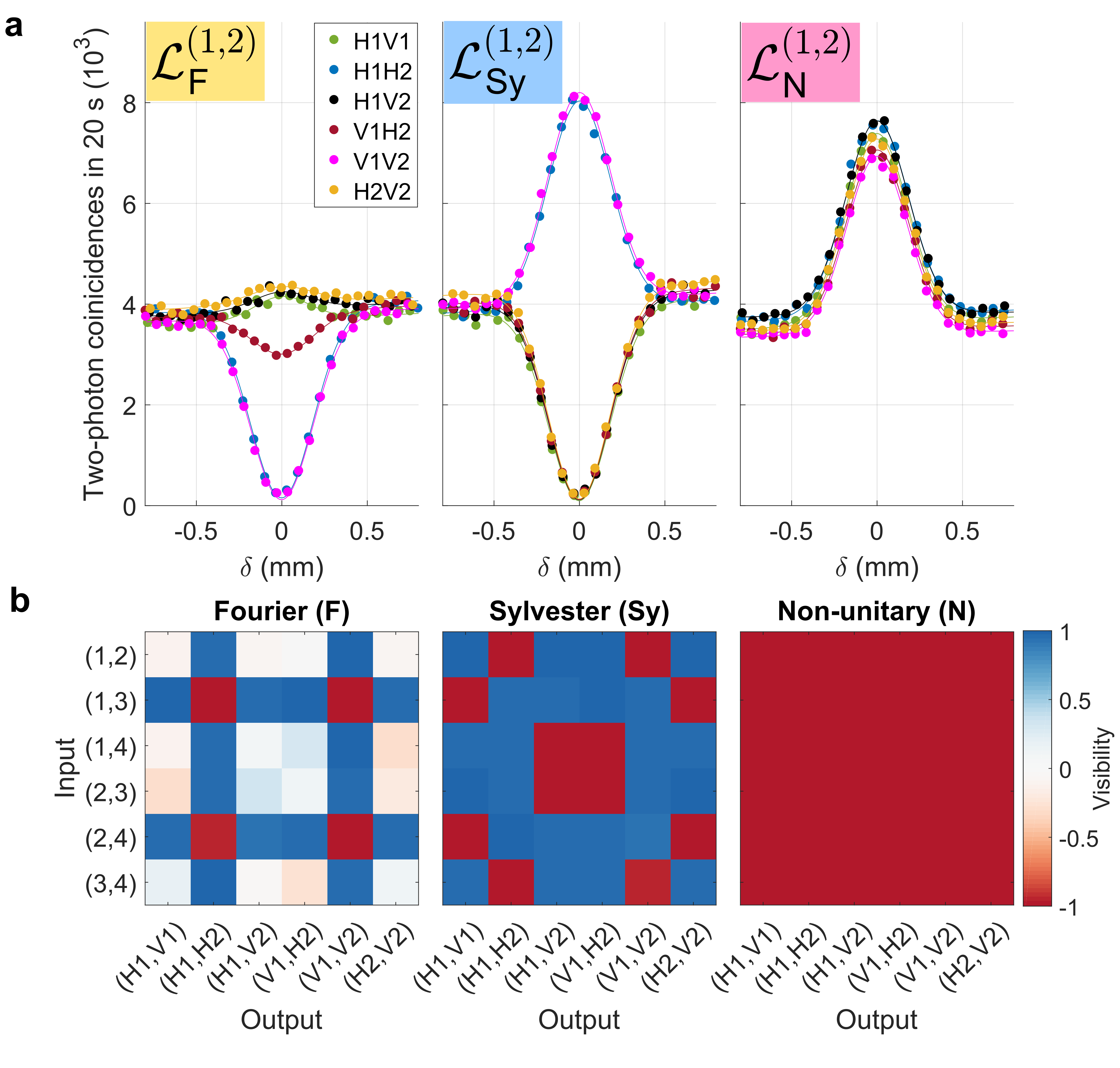}
\caption{\textbf{Control of two-photon interference among spatial-polarisation degrees of freedom.} (a) Two-photon interference: fitting (solid lines) and experiment (dots) for Fourier $\mathcal{L}^{(1,2)}_{\text{F}}$, Sylvester $\mathcal{L}^{(1,2)}_{\text{Sy}}$, and non-unitary $\mathcal{L}^{(1,2)}_{\text{N}}$ transformations where the two-photon state is coupled to the (1,2) input pair. (b) Visibility pattern of four-dimensional Fourier (F), Sylvester (Sy) and non-unitary (N) transformation for all input-output combinations. This corresponds to 18 balanced 4x2 optical networks with fully controllable phase relations.}
\label{fig2}
\end{figure}

We demonstrate deterministic manipulation of two-photon interference through a designed optical network $\mathcal{L}_i$. First, we generate a two-photon state by spontaneous parametric down-conversion (SPDC) process (see Methods) and guide it to the experimental platform (Fig.~\ref{fig1}), in which an optical network $\mathcal{L}_i$ is encoded using the spatial light modulators (SLMs). We implement 4-output $\times$ 2-input optical networks simulating the action of four-dimensional Fourier~\cite{Crespi2016} and Sylvester~\cite{Viggianiello2018} interferometers (see SI Section 4 for definitions). These interferometers are used for certifying indistinguishability between input photons via verifying a suppression criteria~\cite{Tichy2014,Dittel2018}. Here, we verify this criteria for a specific two-photon input state by measuring the full set of output two-fold coincidences (Fig.~\ref{fig2}). Maximum two-photon visibility values measured after propagating through the MMF ($0.96\pm0.01$) and directly at the SPDC source ($0.95\pm0.03$) are the same, showing that the platform does not introduce significant temporal distinguishability between photon pairs (see SI Section 4). The results show quantum distinctive features: values of the degree of violation $\mathcal{D}$, defined as the probability of occupying two-photon states in all suppression configurations~\cite{Crespi2016,Viggianiello2018}, are as small as $0.022\pm0.009$ (the Fourier interferometer, for $(1,3)$ and $(2,4)$ input pairs) and $0.014\pm0.008$ (the Sylvester interferometer, for all input pairs).

Owing to the high number of propagation modes supported by the MMF, we can manipulate phase and amplitude of each element in an optical network independently. To demonstrate this ability, we implement the non-unitary transformation $\mathcal{L}_{\text{N}}$, defined as $\begin{psmallmatrix*}[r]
     1 & -1 \\
     -1 & 1
\end{psmallmatrix*}^{\otimes2}$, which maps all two-photon interferences into photon anti-coalescences (Fig.~\ref{fig2}). The phenomenon presents a distinct result originating from non-unitarity, which derives from information losses stemming from the fact that we do not control all input modes of the MMF. The error between the experimentally synthesised transformation and the theoretically desired one is quantified by $\Delta\mathbb{V}=\langle |V^{\text{exp}}_{ij} - V^{\text{th}}_{ij}|\rangle_{ij}$, where $V^{\text{exp}(\text{th})}_{ij}$ is the experimental (theoretical) visibility over the ($i,j$) output ports. We measure $\Delta\mathbb{V}=0.05\pm0.04$ on average over all transformations (SI Section 4), thus demonstrating accurate control over $4\times2$ linear transformations across spatial-polarisation degrees of freedom.

We now illustrate the use of our experimental platform to simulate coherent absorption, an intriguing phenomenon in quantum transport~\cite{Roger2016}. A typical case is the effect of a lossy beamsplitter on a two-photon N00N state $(\ket{2,0}+e^{2i\phi}\ket{0,2})/\sqrt{2}$. This produces a two-photon absorption probability that depends on the phase $\phi$. The phenomenon has been recently demonstrated using a bulk-optics setup with an absorptive graphene layer~\cite{Roger2016} and a plasmonic metamaterial~\cite{Vest2018,Lyons2019}.
\begin{figure}[htp]
\centering
\includegraphics[width=\linewidth]{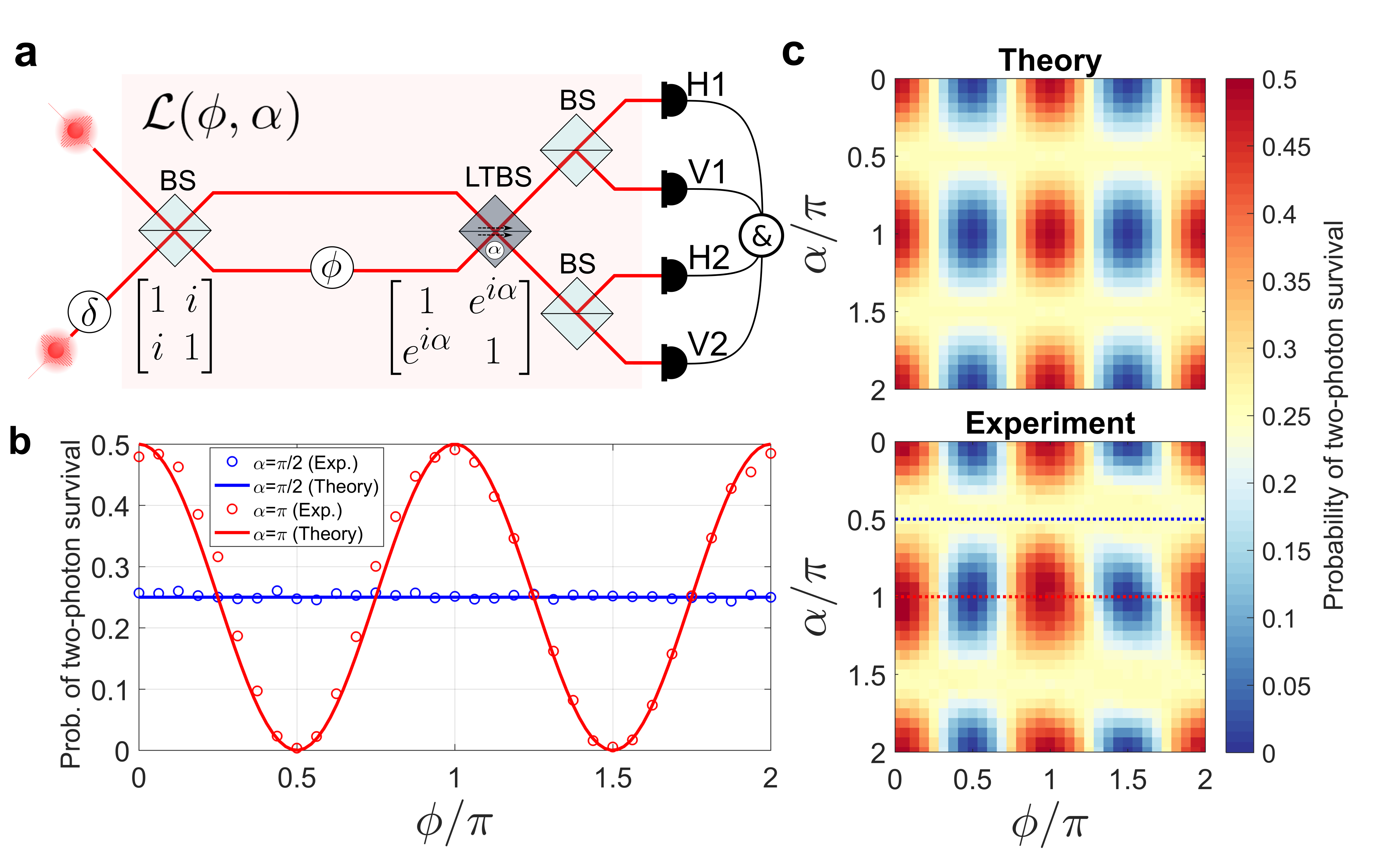}
\caption{\textbf{Controlled coherent absorption.} (a) The linear network $\mathcal{L}(\phi,\alpha)$ programmed in the MMF (Fig.\ref{fig1}) emulates the following circuit: photon pairs enter a Mach-Zehnder (MZ) interferometer composed of a balanced beamsplitter and an LTBS. Both the phase $\phi$ between the two arms and the phase $\alpha$ of the LTBS can be tuned at will. Light in each output port of the MZ interferometer is analysed via two balanced beamsplitters preceding an array of four photocounters to measure the probability of two-photon survival at the targeted output ports. (b) Probability of two-photon survival at the targeted outputs: theory (solid lines) and experiment (dots). The blue dots are for $\alpha=\pi/2$, corresponding to an emulated lossless MZ interferometer. The corresponding probability of two-photon survival is independent of $\phi$. The red dots are for $\alpha=\pi$, corresponding to a lossy beamsplitter in which the probability of two-photon survival depends on the relative phase $\phi$. (c) Probability of two-photon survival as a function of $\phi$ and $\alpha$, showing a transition from emulated lossless to lossy LTBS.}
\label{fig3}
\end{figure}

In our work, we use our fibre platform to simulate the coherent absorption experiment (Fig.~\ref{fig3}a), where the transformation $\mathcal{L}(\phi,\alpha)$ can be seen as a succession of three linear operations: (i) indistinguishable photons are sent onto a beamsplitter to generate a N00N state (N=2) with a controllable output phase $\phi$; (ii) the N00N state interacts with a lossy phase-tunable beamsplitter (LTBS). The matrix that describes the action of the LTBS is $ t\left( \begin{smallmatrix}1 & e^{i\alpha}\\ e^{i\alpha} & 1 \\ \end{smallmatrix}\right)$ where $t\leq 0.5$ is the transmission coefficient and $\alpha$ is a fully tunable phase~\cite{Roger2016}; (iii) the two output ports of the LTBS are distributed into four output ports by two balanced beamsplitters in order to measure two-photon survival probability. This overall survival probability is defined as a sum of probabilities of detecting two photons in all possible output combinations of the LTBS, i.e., both photons on either ports (up or down) or one photon at each port.

As shown in Fig.~\ref{fig3}b, the effect of coherent absorption is maximised for $\alpha=p\pi, p \in \mathbb{Z}$ (red line). In the case where the relative phase $\phi=q\pi, q \in \mathbb{Z}$, which corresponds to having a state $(\ket{2,0}+\ket{0,2})/\sqrt{2}$ as input, the output state is a superposition of vacuum- and two-photon state and the probability of one-photon transmitting to the targeted outputs is null. This result hence exhibits the nonlinear behaviour of the two-photon absorption in the quantum regime. On the other hand, when $\phi=q\pi+\pi/2$, thus corresponding to an input state $(\ket{2,0}-\ket{0,2})/\sqrt{2}$, only single-photon loss occurs
(see SI Section 5 for details). Owing to our ability of fully control the relative phase $\alpha$ (Fig.~\ref{fig3}c), which is a significant step forward with respect to previous experimental arrangements~\cite{Roger2016,Vest2018,Lyons2019}, we observe a transition of the coherent absorption phenomenon from unitary $\alpha=q\pi+\pi/2$ (blue dots) to the maximal coherent absorption situation $\alpha=\pi$ (red dots). 

Partial control, which is usually deleterious for a quantum system, here provides the ability to coherently control the interaction in a non-unitary way, which can be exploited for processing tasks~\cite{Xomalis2018}. Note that, as the optical system (SLM and MMF) is nearly lossless, and non-unitarity in our experiment originates from the fact that we control only half of the propagation modes of the MMF in each input port (see SI Section 3 for explanation). The unmonitored modes thus embody a sink where information about the desired optical network leaks, resulting in effective open system dynamics of the latter. The total energy transmittance $2\vert t\vert^2$ to all targeted outputs of the optical network $\mathcal{L}_i$ reaches 0.45(0.5) experimentally(theoretically), which is close to the maximum theoretical transmission of the LTBS.

\begin{figure}[htp]
\centering
\includegraphics[width=\linewidth]{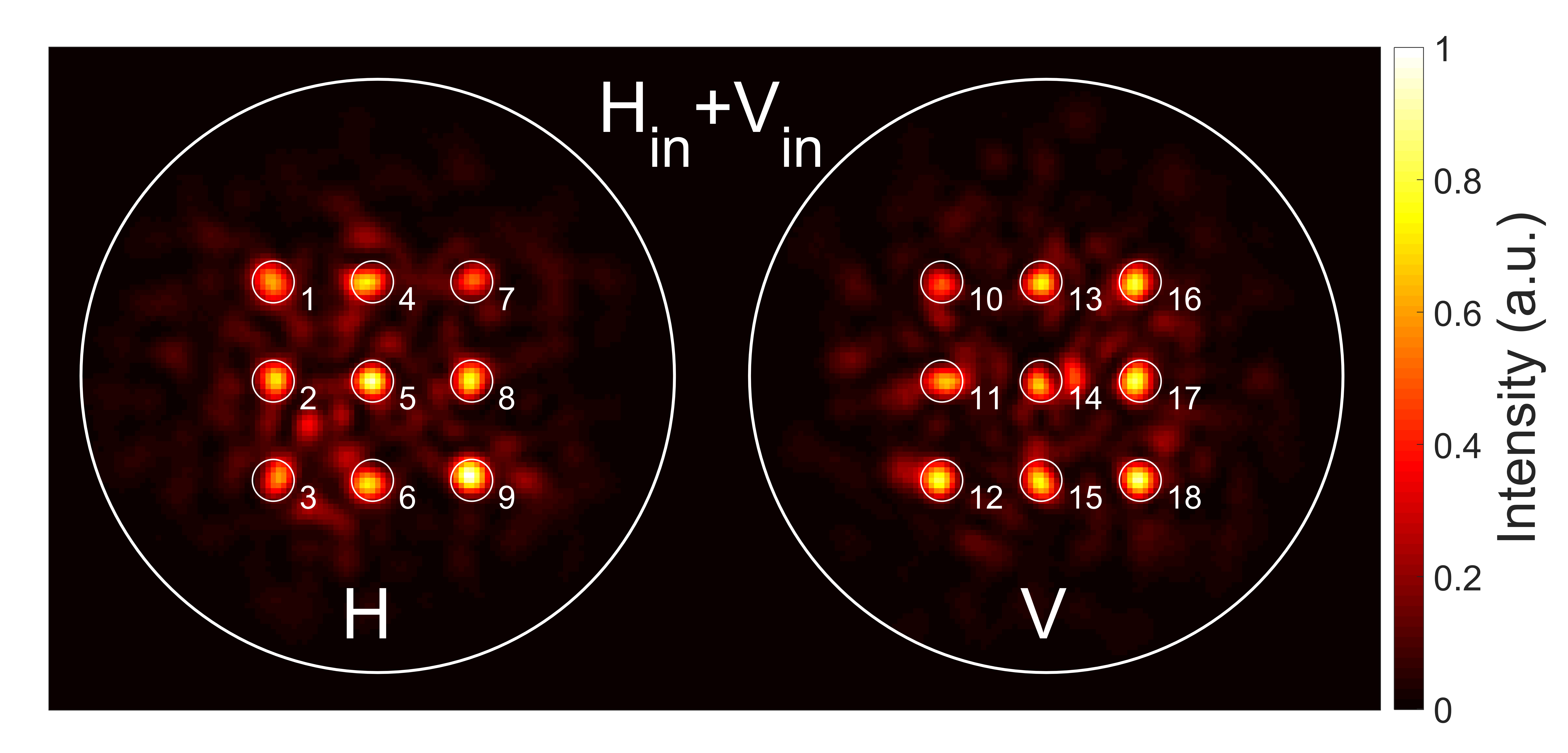}
\caption{\textbf{Intensity image of a high-dimensional linear-optical network on the EMCCD.} The SPDC light from both inputs is simultaneously distributed into 18 targeted outputs, 9 in each polarisation (H: Horizontal; V: Vertical).}
\label{fig4}
\end{figure}

The dimensionality of our platform can in principle be scaled up, as the main limiting factor in our experimental implementation is given by the detection architecture. A significantly larger network can be managed, for instance, by replacing our detection apparatus with an array of correlation detectors~\cite{Defienne2018b}. In Fig.~\ref{fig4}, we experimentally showcase the scalability of our platform by designing a larger optical network with 18 targeted outputs allocated arbitrarily at different positions and with arbitrary polarisation on the EMCCD camera. In SI Section 3, we discuss the fidelity, scalability and programmability of this optical network architecture.

We report the use of a multimode fibre to implement fully programmable linear optical networks across spatial and polarisation degrees of freedom. This platform harnesses the highly complex coupling between a large number of modes of the MMF, thanks to the ability to spatially control the input light wavefront. We successfully programmed this platform to implement circuits able to tackle certification tasks all the way up to the emulation of coherent absorption. We thus demonstrate the versatility and full reconfigurability of our approach, including the management of different degrees of freedom of the propagating light.
Complex mixing occurring in an optical mixer, in general, can go beyond path and polarisation reported in this work. Spectral, temporal, and spatial (radial and orbital angular momentum) degrees of freedom can also be manipulated~\cite{Rotter2017,Fickler2017}. We anticipate that our architecture can be applied to those degrees of freedom. We also highlight its scaling potential by demonstrating control over up to 18 output ports, whereas the number of input ports can also be scaled well beyond 2, provided a multi-photon source is available. Our architecture provides an efficient and scalable alternative to integrated circuits for linear quantum networks.

\section*{Methods}
\textbf{Two-photon source.} The frequency-degenerate photon pairs are produced from a type-II polarisation-separable collinear spontaneous parametric down-conversion (SPDC) source (Fig.~\ref{fig1}a), using a 10-mm periodically poled potassium titanyl phosphate crystal (ppKTP) pumped by a single-mode continuous-wave laser in a single spatial mode configuration. The photon pairs transmit through a spectral filter ($\lambda=810\pm5$ nm) and are separated by a polarising beamsplitter. The indistinguishability of photon pairs is controlled by a temporal delay $\delta$. The photon pairs are then prepared in the same horizontal polarisation, and collected with polarisation-maintaining single-mode fibres, which are then connected to the MMF platform. A coincidence window was set at 2.5 ns for all experiments. All coincidence counts are corrected for accidental coincidence counts. 

\textbf{Network programming.} After the TM acquisition using a phase-shifting holographic technique with a co-propagating reference~\cite{Cizmar2011} (see SI Section 1), a given linear transformation $\mathcal{L}_i$ (network) is programmed. The input electric fields $\tilde{E}_{\text{in}}^{(j)}$ and the corresponding SLM phase patterns for each $j$-th input port is calculated by solving an inverse scattering problem $\tilde{E}_{\text{in}}^{(j)}=\mathrm{T}^{(j)\dagger} \mathcal{L}^{(j)}_{i}$, where $\mathrm{T}^{(j)}$ is the sub-part of the measured TM linking the relevant input modes for each $j$-th input port to the targeted output modes. Imperfections in generating the input electric fields $\tilde{E}_{\text{in}}$ with the SLM lead to errors in the coefficients of the linear transformation $\mathcal{L}_i$. In the case of our first experiment (control of two-photon interference), we additionally performed an amplitude correction when a new $\mathcal{L}_i$ is programmed by adjusting on the amplitudes of the co-propagating reference fields. This was done by minimising the mean squared error between implemented amplitudes and the desired ones. For the experiment on the control of the coherent absorption, we compensated the amplitude variations using the normalised second-order correlation function $g^{(2)}$.

\section*{Data Availability Statement}
The data that support the plots within this paper and other findings of this study are available from the corresponding author upon reasonable request.

\section*{Code Availability Statement}
The code for data analysis and simulation that support the plots within this paper and other findings of this study are available from the corresponding author upon reasonable request.

\section*{Competing financial interests}
The authors declare no competing financial interests.


\section*{Acknowledgements}
 The authors thank C. Moretti for technical support. The work is supported by European Research Council (ERC) (724473). S.G. is a member of the institut universitaire de France (IUF). M.P. is supported by the European Commission through the H2020 Collaborative project `Testing the large-scale limit of quantum mechanics' (TEQ, grant number 766900), the Science Foundation Ireland-Department for Economy Investigator Programme `Quantum control of nanostructures for quantum networking' (QuNaNet, grant number 15/IA/2864), the Leverhulme Trust through the Research Project Grant `Ultracold quantum thermo-machine' (UltraQuTe, grant number RGP-2018-266), MSCA co-funding of regional, national and international programmes (grant number 754507), and COST Action CA15220 `Quantum Technologies in Space (QTSpace)'. L.I. acknowledges partial support from Fondazione Angelo Della Riccia. T.J. was supported by an Human Frontier Science Program Cross-Disciplinary Fellowship (LT000345/2016-C), and the ERC (758752). S.L. acknowledges support from a Franco-Thai Scholarship. 

\section*{Statement of author contribution}
S.L., T.J., H.D. carried out the experiment and the analysis of the data, S.L. L.I. performed numerical simulations and L.I., A.F., M.P. provided a theoretical analysis of the results. S.L. proposed the coherent absorption experiment. S.G. proposed the original idea and supervised the project. All authors discussed the implementation, the experimental data and the results. All authors contributed to writing the paper.
\onecolumngrid
\hspace{3cm}
\newpage
\begin{center}
\textbf{\large Supplementary information for: \\ Programmable linear quantum networks with a multimode fibre}\\[1cm]
\end{center}
\renewcommand\thesection{\arabic{section}}
\setcounter{equation}{0}
\setcounter{figure}{0}
\setcounter{table}{0}
\setcounter{page}{1}
\renewcommand{\theequation}{S\arabic{equation}}
\renewcommand{\thefigure}{S\arabic{figure}}

\section{Acquisition of the transmission matrix}
\label{SI_Acquisition of the transmission matrix}
\begin{figure*}[htp]
\centering
\includegraphics[width=0.95\linewidth]{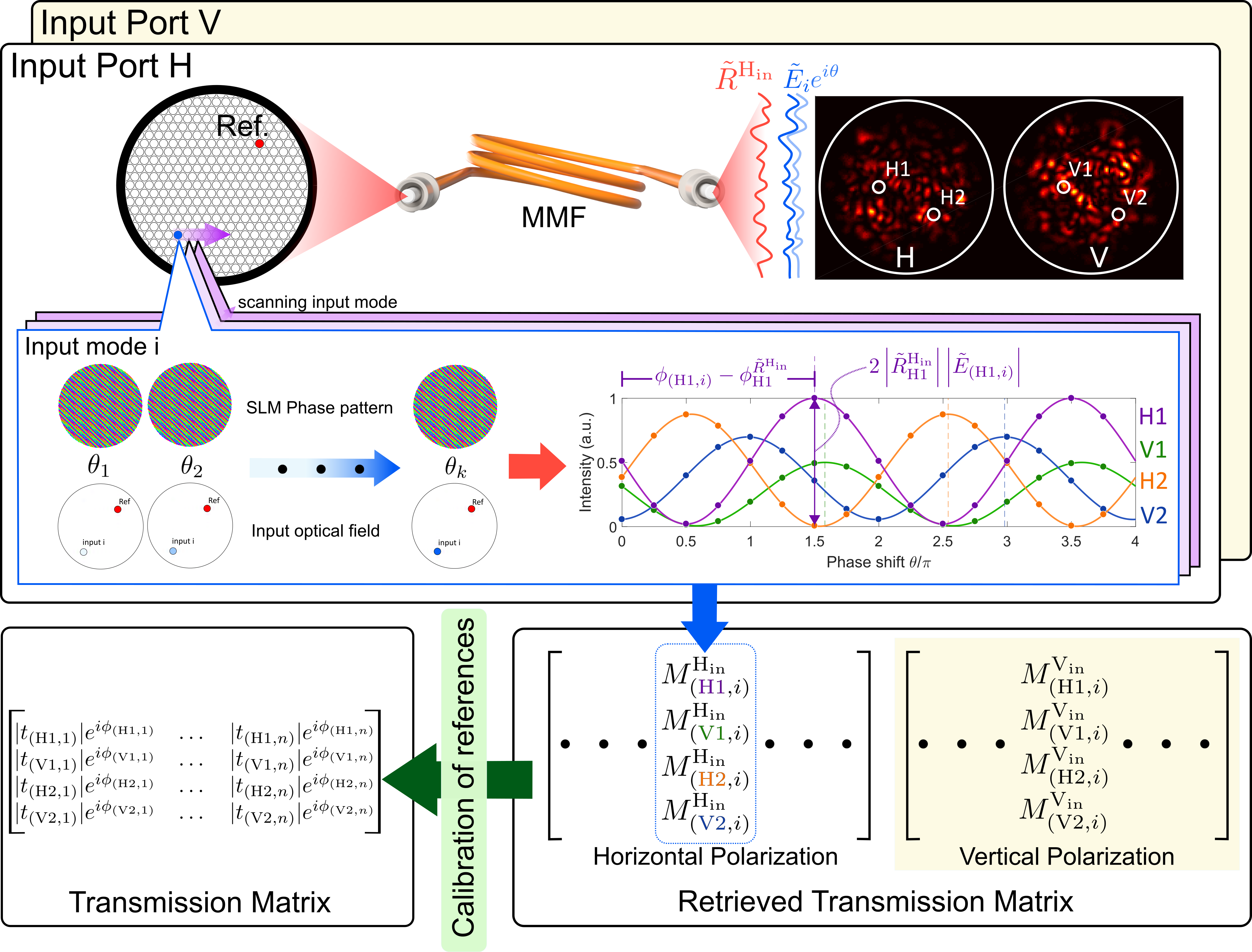}
\caption{Acquisition of the transmission matrix.}
\label{figS0}
\end{figure*}

The Transmission Matrix (TM) of a graded-index Multimode Fibre (MMF) (Thorlabs, GIF50C: of length of 55.3$\pm$0.1 cm, core diameter of 50$\pm$2.5 $\mu$m, and numerical aperture of 0.200$\pm$0.015) is acquired using a phase-shifting holographic technique with a co-propagating reference~\cite{Cizmar2011,Rotter2017}. For each input port, an over-complete spatial basis set of input wavefronts is sequentially sent through the MMF using the Spontaneous Parametric Down-Conversion (SPDC) source. The input basis is defined as a set of focus spots on an isometric grid on the input face of the MMF as shown in Fig.~\ref{figS0}. By shifting the phase of each input mode relative to the co-propagating reference, the amplitude and phase of all targeted outputs are retrieved simultaneously using avalanche photodiodes (Similarly, all outputs can also be retrieved using an Electron-Multiplying CCD (EMCCD) camera for analysis in Sec.~\ref{SI_Transmission Matrix and its Properties}). The part of TM for each input port is independently acquired. Consecutive measurements for different input ports require a calibration of the relative amplitudes and phases of the co-propagating reference fields. In order to do so, we program a linear transformation $\mathcal{L} \propto \bigl( \begin{smallmatrix}1 & 1 & 1 & 1\\ 1 & 1 & 1 & 1\\ \end{smallmatrix}\bigr)^{\intercal}$ using the pre-calibrated TM (see Network programming in Methods). We then use the corresponding measured photon counts and two-photon coincidences to obtain an initial value of the relative co-propagating reference fields (see Ref.~\cite{Laing} for details). We further calibrate the reference field by minimizing $\Delta\mathbb{V}$, where $\Delta\mathbb{V}=\langle |V^{\text{exp}}_{ij} - V^{\text{th}}_{ij}|\rangle_{ij}$ and $V^{\text{exp}(\text{th})}_{ij}$ is the experimental (theoretical) visibility of two-photon interference at the $(i,j)$ pair of detectors.
\newpage

\twocolumngrid
\section{Properties of the Transmission Matrix of Multi-Mode Fibres}
\label{SI_Transmission Matrix and its Properties}
\begin{figure}[htp]
\label{figS1}
\centering
\includegraphics[width=\linewidth]{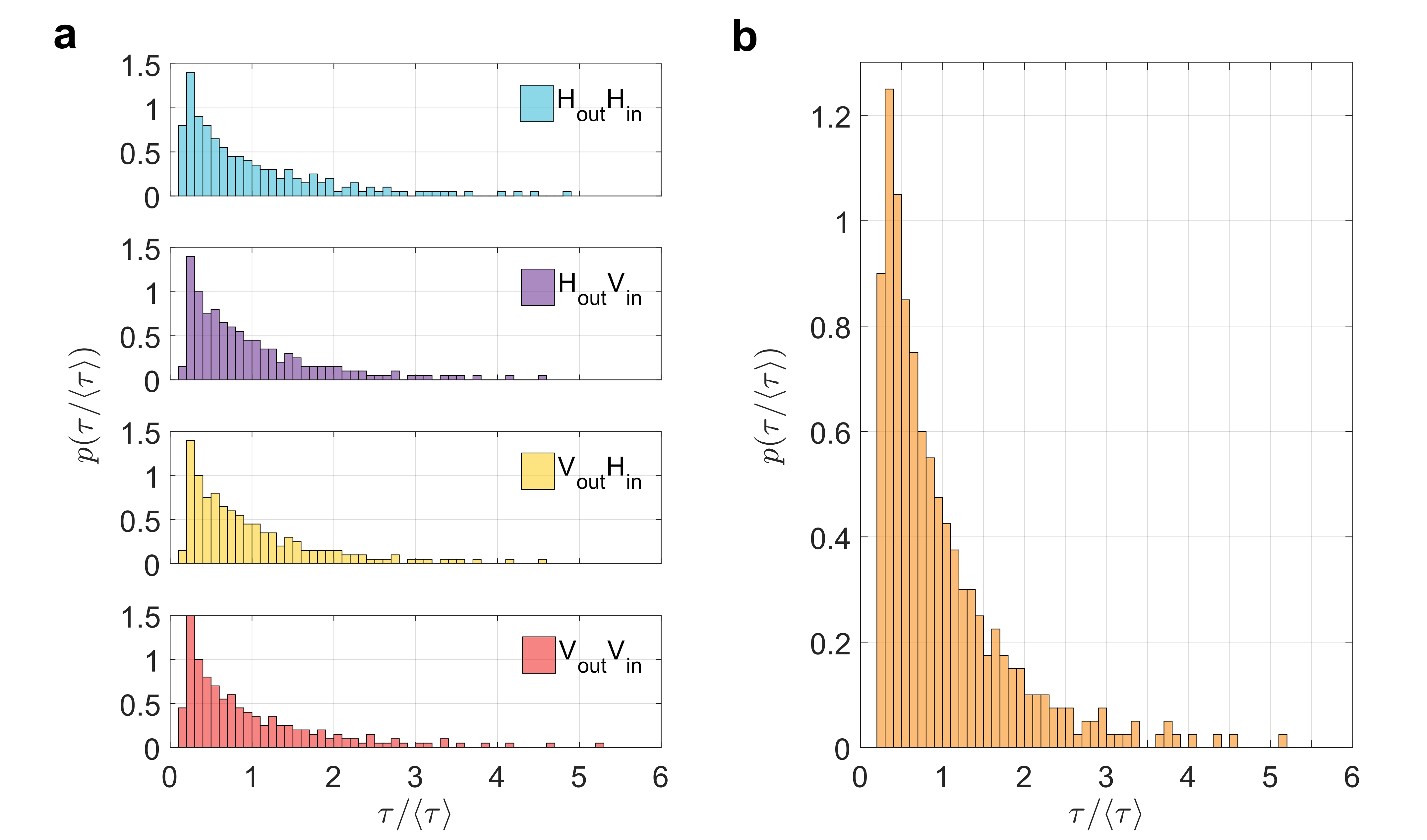}
\caption{(a) Probability distribution of transmission eigenvalues $p(\tau/\langle\tau\rangle)$ for each polarisation channel (H$_{\text{out}}$H$_{\text{in}}$, H$_{\text{out}}$V$_{\text{in}}$, V$_{\text{out}}$H$_{\text{in}}$, V$_{\text{out}}$V$_{\text{in}}$) (b) Probability distribution of transmission eigenvalues $p(\tau/\langle\tau\rangle)$.}
\end{figure}

It has recently been shown that the MMF used here have a refractive index profile that deviates from a perfect parabola, thus presenting mode coupling between Propagation-Invariant Modes~\cite{Flaes2018}. As a consequence, the speckle pattern appearing after propagation along the MMF results both from the phase delays between modes of the fibre and from the mode coupling due to imperfections of the refractive index profile along the propagation axis, due for instance to bending and twisting of the fibre itself. The TM of the MMF is thus expected to induce significant mixing across modes, irrespectively of the basis being used. This implies that any targeted output mode can be excited by injecting combinations of many spatial and polarisation modes. For instance, in order to show the polarisation mixing, we experimentally measure the probability distribution of transmission eigenvalues $\tau$ \footnote{The transmission eigenvalues $\tau$ are obtained from singular value decomposition of $\mathrm{T}^{\dagger}\mathrm{T}$, where $\mathrm{T}$ is a part of TM of interest.} for the part of the full TM, noted as $\mathrm{T}$, corresponding to each input-output polarisation channel (H$_{\text{out}}$H$_{\text{in}}$, H$_{\text{out}}$V$_{\text{in}}$, V$_{\text{out}}$H$_{\text{in}}$, V$_{\text{out}}$V$_{\text{in}}$), and observed a similar distribution for all polarisation pairs (Fig.~\ref{figS1}a). Similarly, the overall probability distribution of the transmission eigenvalues $\tau$ of the full TM has been investigated experimentally (Fig.~\ref{figS1}b). We have found that this distribution can be described by a model based on random-matrix theory recently proposed in Ref.~\cite{Chiarawongse} that is based on the free probability theory and the Filtered Random Matrix ensemble. This is also verified by checking that it is possible to focus on any spatial and polarisation state of the output plane (within the fibre core) with high efficiency, while keeping a low unstructured background (data not shown).

The complex mixing induced by a MMF has recently allowed to achieve complete control on a polarisation state of the output field, by modulating only the spatial profile in a graded-index fibre~\cite{Xiong2018} and programming a linear transformation on a spatial degree of freedom in a step-index fibre via the optimisation approach proposed in~\cite{Matthes2018}. The ability to design a linear transformation based on complex mixing finds use also in different scenarios, for example, in the context of wireless communications in a controllable indoor environment~\cite{DelHougne2018}.

\section{Reliability, scalability and programmability of the optical network}
\label{sec:Fidelity, scalability and programmability of the optical network}

We now study the usability of our method to faithfully reproduce a given target linear evolution $\mathcal L$, modelled as a $k \times m$ matrix, where $m$ ($k$) is the number of input (output) ports of the associated optical network. We define $n$ as the number of propagation modes of the MMF. We assume that the Spatial Light Modulator (SLM) provides complete control over all $n$ propagation modes of the MMF. Thus, the number of tunable elements in the setup is also $n$, and for each of the $m$ input ports of the target optical network, we can control and inject $n/m$ input physical modes of the MMF.

First, we quantify the ability to theoretically program a linear transformation $\mathcal L$. Let us first denote with $\mathrm{T}^{(j)}$ the part of transmission matrix linking the $j$-th input port to the $k$ output ports of interest, which is thus a $k\times(n/m)$ matrix. It is important to note that here with ``j-th input por'' we mean a set of $n/m$ physical modes controlled by a given SLM, which are used collectively to reproduce the action of the target evolution $\mathcal L$ on the $j$-th input port. The corresponding column of $\mathcal L$ is thus denoted as $\mathcal{L}^{(j)}$. The input optical field $\tilde{E}_{\text{in}}^{(j)}$ that reproduces the target evolution $\mathcal{L}^{(j)}$ is determined by solving the relation $\tilde{E}_{\text{in}}^{(j)}=\mathrm{T}^{(j)\dagger}\mathcal{L}^{(j)}$. With this notation, $\tilde{E}_{\text{in}}^{(j)}$ is thus a vector of length $n/m$, which represents the set of amplitudes of the input wavefront that, when displayed on the SLM, results in an effective implemented set of output amplitudes, here denoted as $\tilde{\mathcal{L}}^{(j)}$. Then, the corresponding phase patterns for all input ports are displayed on the SLM and light propagates through the fibre. MMF and SLMs thus work together to implement an effective linear optical network, which we describe with the matrix $\tilde{\mathcal{L}}$, that corresponds to desired $\mathcal{L}$ up to a global amplitude and phase factor. For each $j$-th input port, $\tilde{\mathcal{L}}^{(j)}$ and $\mathcal{L}^{(j)}$ are related via
\begin{equation}
\label{eq:Mapping}
    \tilde{\mathcal{L}}^{(j)}=\mathrm{T}^{(j)}\mathrm{T}^{(j)\dagger}\mathcal{L}^{(j)}.
\end{equation}

Being $\mathrm{T}^{(j)}$ a $k \times n/m$ matrix, $\mathrm{T}^{(j)}\mathrm{T}^{(j)\dagger}$ is a $k \times k$ matrix. The overall fidelity of the optical network can therefore be related to the so-called \textit{time reversal operator}, $\mathrm{T}\mathrm{T}^\dagger$~\cite{Popoff2010}, which is in general close to the identity for random matrices with independent and identically distributed (i.i.d.) entries.

We quantify the difference between implemented and desired evolution using the quantity $\mathcal F$, defined as: 
\begin{equation}
\label{eq:Fidelity}
    \mathcal{F}(\tilde{\mathcal{L}},\mathcal{L}) = 1 -  \frac{\|\mathcal{L} - \tilde{\mathcal{L}}\|}{mk},
\end{equation}
where $\|\cdot\|$ is the $l_{1}$-vector norm, defined as $\|A\|=\sum_{i=1}^m\sum_{j=1}^k |a_{ij}|$.
The last term in Eq.~\eqref{eq:Fidelity} measures an average element-wise distance between $\mathcal{L}$ and $\tilde{\mathcal{L}}$. In the following, we refer to $\mathcal F$ as the figure of merit (fidelity) to characterise the performance of our implementation.

We will now evaluate $\mathcal F$ in three different cases: modelling $\mathrm T$ as a Random Matrix (RM), modelling $\mathrm T$ as a Random \textit{Unitary} Matrix (RUM), and directly using the experimentally measured TM (MMF). In the RM case, $\mathrm T$ is taken to be an $n\times n$ complex matrix whose components are sampled from the i.i.d. complex Gaussian ensemble. In the RUM case, we take $\mathrm T$ to be a random unitary matrix, generated via orthogonal triangular decomposition of a RM. Finally, we also compare these two cases with the experimentally measured TM (MMF).

When $\mathrm T$ is RM, the operator $\mathrm T\mathrm T^\dagger$ can be estimated explicitly~\cite{Derode2001,Aubry2010a}, and shown to equal

\begin{equation}
\label{eq:TT'}
    \mathrm{T}\mathrm{T}^\dagger=\mathrm{I}+\mathrm{H}/\sqrt{n},
\end{equation}
where $\mathrm H$ is a complex Hermitian noise matrix. $\mathrm T\mathrm T^\dagger$ clearly converges to the identity operator $\mathrm{I}$ with an additional term whose root-means-square amplitude scales as $1/\sqrt{n}$. Similarly, when only $n/m$ input modes are controlled for each input port, the corresponding time reversal operator $\mathrm{T}^{(j)}\mathrm{T}^{(j)\dagger}$ converges to the identity operator with an additional term scaling as $\sqrt{m/n}$. By substituting $\mathrm{T}^{(j)}\mathrm{T}^{(j)\dagger}$ into Eq.~\ref{eq:Mapping} and then $\tilde{\mathcal{L}}$ into the definition of the $\mathcal F$ in Eq.~\ref{eq:Fidelity}, we find that the fidelity scales as
\begin{equation}
\label{eq:FidelityScaling}
    \mathcal{F}(\tilde{\mathcal{L}},\mathcal{L})=1-\mathcal{O}\left(\sqrt{\frac{m k}{n}}\right).
\end{equation}
The RM model, albeit a simple one, provides strong evidence in support of the statement that arbitrary desired transformations can be implemented with high fidelity, even when the dimension of the problem is scaled up.

\begin{figure}[htp]
\centering
\includegraphics[width=\linewidth]{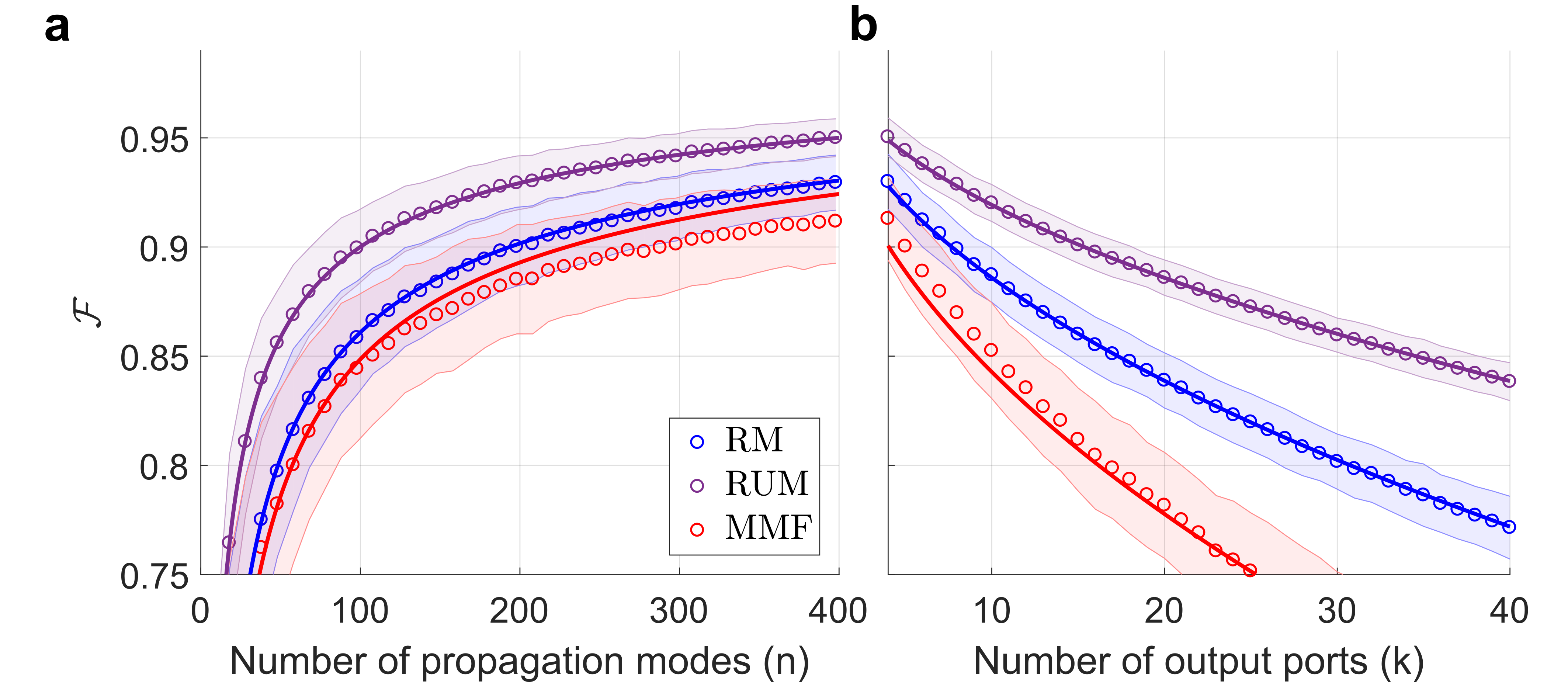}
\caption{(a) Fidelity $\mathcal{F}$ of an optical network ($m=2$, $k=4$) as a function of the number $n$ of propagation modes supported by a medium (b) as a function of the number $k$ of the targeted outputs ($m=2$, $n=398$). The mean (circle) and standard deviation (shaded area) of the fidelity are calculated from simulating 1000 $m \times k$ linear networks arbitrarily defined on different targeted outputs of a high-dimensional $n \times n$ random matrix (RM, RUM and MMF). The $m \times k$ desired linear networks are generated randomly (i.e., elements are i.i.d. complex Gaussian coefficients). The high-dimensional $n \times n$ random matrix are generated either from a random matrix (RM, blue circle), a random unitary matrix (RUM, purple circle), or the experimentally measured transmission matrix (MMF, red circle). In the case of the MMF, we reduce the number of the propagation modes $n$ by randomly selecting $n$ columns and rows of the measured full TM. All curves show the predicted $1-\mathcal{O}(\sqrt{mk/n})$ behaviour.}
\label{figS2}
\end{figure}

Furthermore, we study numerically the behaviour of the fidelity $\mathcal F$ with which arbitrary desired transformations are obtained, using the three different TM models introduced previously (Fig.~\ref{figS2}). In our numerical model, in order to account for the noise on the other output modes, we define $\mathcal{L}$ on an output space of dimension $n$ where the $(n-k)$ rows of $\mathcal{L}$ corresponding to unassigned output modes are set to zero. As shown in Fig.~\ref{figS2}a, for optical networks of dimension $4 \times 2$, the fidelity $\mathcal{F}$ scales as expected as $1-\mathcal{O}(1/\sqrt{n})$ when we increase the number of modes of the complex medium. For a fixed $n$, the fidelity decreases when increasing the number $k$ of targeted output ports, following $1-\mathcal{O}(\sqrt{k})$ (Fig.~\ref{figS2}b).
For both graphs, the RUM provides the highest fidelity since it ensures energy conservation \footnote{RUM provides the highest fidelity when the number $m$ of input ports is low. As $m$ increases, the fidelity converges to the value provided by the RM because the unitary condition relieves}, while the fidelity with the MMF model is slightly below the RM one.  This could be due to mesoscopic correlations~\cite{Rotter2017}, the variation of the enhancement at different targeted outputs due to the co-propagating speckle reference \cite{Cizmar2011}, and to mode-dependent loss~\cite{Carpenter2014a,Chiarawongse}.

Following Eq.~\ref{eq:FidelityScaling}, a high degree of programmability and high fidelity are achieved when $m,k \ll n$. The programmability of our optical network architecture originates from how well the SLMs control interference, and is made possible by the high-dimensional intermodal coupling provided by the MMF. The two key properties are: (1) The fibre-induced complex mixing across modes and degrees of freedom. This allows to assign $k$ targeted output ports of arbitrary position and polarisation to a corresponding desired complex coefficient of $\mathcal{L}$. (2) The degree of SLM control which is proportional to $n/(mk)$ (as seen in Eq.~\ref{eq:FidelityScaling}). For each column of the desired transformation $\mathcal{L}$, one has a high number of programmable SLM elements $(n/m)$ and corresponding complex MMF intermodal coupling coefficients. Increasing the number of targeted output modes $k$, on the other hand, reduces the degree of SLM control. This effect experimentally corresponds to a cross-talk between targeted outputs due to an overlap of speckle background fields.

In the following we compare our programmable platform to a conventional architecture, in which a large number of phase shifters in cascaded Mach-Zehnder interferometers has to be controlled~\cite{PhysRevLett.73.58}. While the conventional architecture follows a “bottom-up” approach, where a large desired transformation is constructed from a series of small $2\times 2$ ones, our platform can be considered to follow a “top-down” approach, in that the desired transformation is obtained directly, without the need to decompose it in terms of simpler components. In a conventional architecture, to have a fully reconfigurable $k\times k$ unitary transformation $\mathcal{O}(k^2)$ tunable optical elements are required~\cite{PhysRevLett.73.58,Miller2013,Clements:16,Tillmann2016,Tischler2018}.
In our setup, on the other hand, since we have $n$ tunable elements at the input of the complex medium, we expect to be able to program unitaries of dimension up to $m=k=\sqrt{n}$. For a given network, scaling to a larger complex medium (larger $n$, which could be realised for instance by increasing the diameter or the numerical aperture of the multimode fibre) allows increasing the fidelity to values close to unity.

An interesting feature of our method is that the overall enhancement of the photon counts does not significantly depend on the number of targeted output modes $k$ (i.e., the number of detectors). This is well-known from the first paper on wavefront shaping through complex media~\cite{Vellekoop2007}, where it was noted that focusing on $k$ target points instead of on a single one results in a $k$-fold reduction of the intensity per target. From a matrix perspective, the total photon flux that can theoretically be focused on a target output mode $i$ is equal to $\sum_{j=1}^{p}|t_{ij}|^2$, where the $t_{ij}$ are the elements of the TM and $p$ is the number of modes that are controlled, that is approximately equal to $n/m$~\cite{Vellekoop2015}. Increasing $k$ is equivalent to a change of basis and does not significantly modify the total intensity. In our experiment, the overall energy transmittance $\gamma$, defined as the ratio of the photon flux carried by the targeted outputs and the total photon flux transmitted through the MMF, can reach 0.45 for each input port, given the fact that we control only half of the number of propagation modes. This allows us to use this platform to emulate the coherent absorption effect close to the critical transmission of 0.5. The state-of-the-art $\gamma$ of 0.6(0.8) was experimentally reported in a step-index MMF for a conserved circular input polarisation (both linear polarisation channels) \cite{Cizmar2011,Ploschner2015,Turtaev2017}.

\section{Control of two-photon interference across degrees of freedom}
\label{sec:SI_Control of two-photon interference across degrees of freedom}

Indistinguishability of photons can be observed by interfering them in a linear network. The degree of indistinguishability can be inferred from the permanent of the respective linear transformation and the visibility of two-photon interference at the outputs. The latter is defined as $V=(P^{(\text{D})}-P^{(\text{I})})/P^{(\text{D})}$, where $P^{(\text{I/D})}$ is the probability in the case of indistinguishable (I) and distinguishable (D) input photon pairs, respectively. In this work, we implement $4\times 2$ optical networks simulating the action of four-dimensional Fourier~\cite{Tichy2010} and Sylvester~\cite{Crespi2015}, and non-unitary interferometers. Here we provide the respective definitions. An 4-dimensional Fourier interferometer is defined as one implementing the unitary transformation $\mathcal{L}_{\text{F}}$ defined element-wise as $(\mathcal{L}_{\text{F}})_{jk}\equiv\exp(i\pi(j-1)(k-1)/2)/2$. A Sylvester interferometer implements the transformation $\mathcal{L}_{\text{Sy}}\equiv H^{\otimes 2}$, with $H$ the $2\times 2$ Hadamard matrix. These two interferometers are useful for the certification of the indistinguishability between input photons~\cite{Tichy2014,Dittel2018a}. The non-unitary transformation $\mathcal{L}_{\text{N}}$ is defined as $\begin{psmallmatrix*}[r]1 & -1\\ -1 & 1 \\ \end{psmallmatrix*}^{\otimes2}$. It provides the photon anti-coalescences occurring in all input-output pairs. This results purely stems from non-Hermitian physics of two-photon interference, and is not related to anti-coalescences typically observed from two-photon interference of a singlet Bell entangled state on a unitary non-polarising beamsplitter.

\begin{figure}[htp]
\centering
\includegraphics[width=\linewidth]{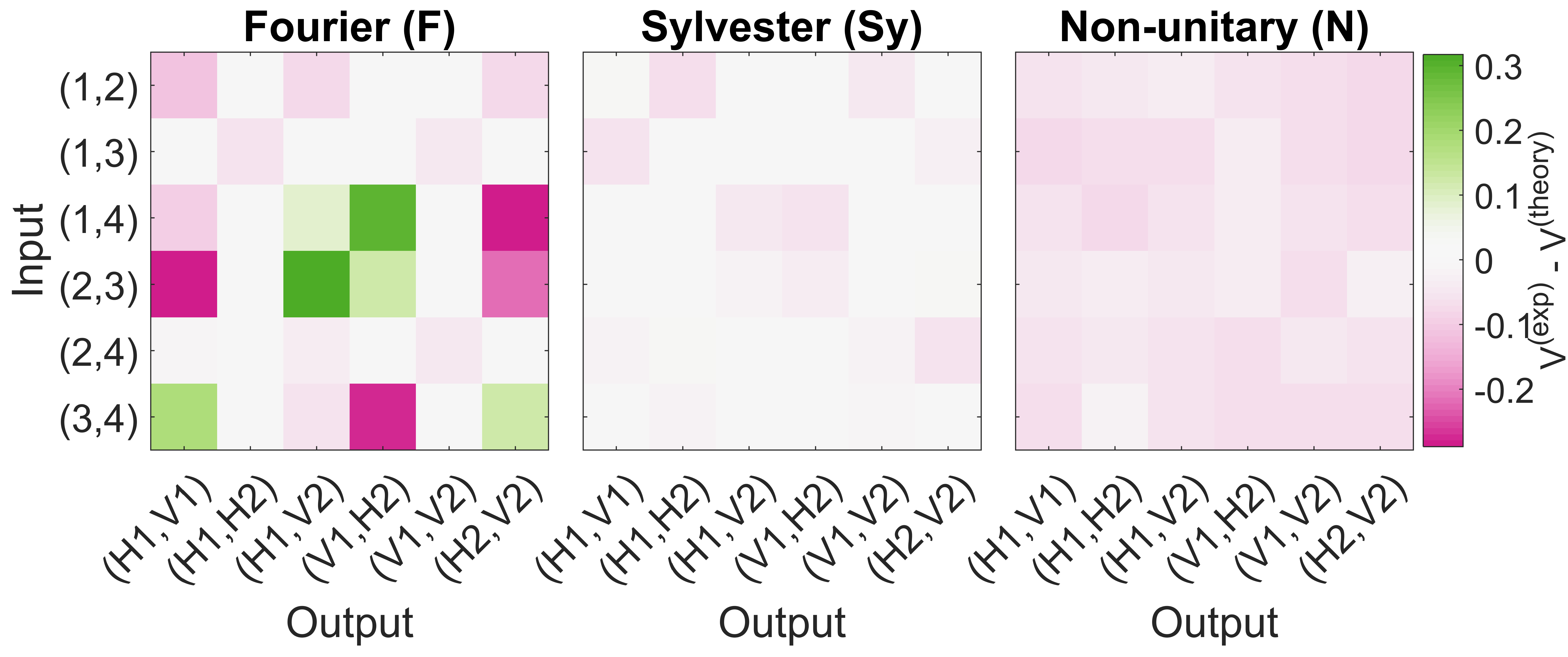}
\caption{Difference between the experimental and theoretical visibility of two-photon interference. We obtained $\Delta\mathbb{V}_{\text{F}}=0.08 \pm 0.06$, $\Delta\mathbb{V}_{\text{Sy}}=0.02 \pm 0.01$, and $\Delta\mathbb{V}_{\text{N}}=0.06 \pm 0.01$ for Fourier, Sylvester, and the non-unitary transformation, respectively.}
\label{figS3}
\end{figure}

We provide a statistical analysis for the experiment on the control of two-photon interference. We determine $\Delta\mathbb{V}=\langle |V^{\text{exp}}_{ij} - V^{\text{th}}_{ij}|\rangle_{ij}$, where $V^{\text{exp}(\text{th})}_{ij}$ is the experimental (theoretical) visibility of two-photon interference at the $(i,j)$ pair of detectors as shown in Fig.\ref{figS3}.

We reconstruct the experimental linear transformations $\tilde{\mathcal{L}}$ with the measured two-photon visibility $V^{\text{exp}}$ by minimizing $\langle |V_{ij} - V^{\text{exp}}_{ij}|\rangle_{ij}$, averaging over the $ij$ pair of detectors. Here we only allow changes in the phase components of a linear transformation $\tilde{\mathcal{L}}$. We obtain the fidelity $\mathcal{F}$ (Eq.\ref{eq:Fidelity}) of $0.95\pm0.03$ (Fourier), $0.98\pm0.01$ (Sylvester), and $0.97\pm0.02$ (Non-unitary), respectively. This is consistent with the numerically calculated fidelity in Fig.~\ref{figS2}.

One potential problem is that the fibre may introduce temporal distinguishability. Experimentally, over all HOM measurements shown in Figure 2, we measured $<1\%$ temporal shift, while the width and visibility of the HOM dips and peaks were also not significantly affected. This could be due to the fact two-photon state experiences nearly the same dispersion effect and may benefit from non-local dispersion cancellation~\cite{Steinberg1992}. This is consistent with the fact that for the fibre we used, the dispersion is 0.26 ps/m, meaning that for our 55 cm-long fibre the maximal dispersion should be on the order of 140 fs (to be compared with 1.5 ps FWHM of the HOM interference).

\section{Theory of coherent absorption}
\label{sec:SI_coherent_absorption}

We now give a brief summary of the theory underlying the coherent absorption presented in Refs.~\cite{Jeffers2000,Roger2016,Vest2018}. In particular, we report on the calculations showing how specific two-photon input states result in a coherent-absorption effect. We refer to Refs.~\cite{Barnett1998,Uppu2016} for a thorough analysis of the lossy beamsplitter operation. We present here a simple model of a lossy beamsplitter based on the use of an auxiliary mode. This allows us to describe a lossy evolution of two modes as a unitary dynamics of three modes, one of which is discarded. The latter can be considered as a transformation evolving each of the two input modes of the device into three output modes as
\begin{equation}
\begin{aligned}
    a_1^\dagger \to u_{11}a_1^\dagger + u_{21}a_2^\dagger + u_{31}a_3^\dagger, \\
    a_2^\dagger \to u_{12}a_1^\dagger + u_{22}a_2^\dagger + u_{32}a_3^\dagger.
\end{aligned}
\end{equation}
Here, $a_1,a_2$ denote the destruction operator for the first two input/output modes of the beamsplitter, while $a_3$ is used to denote an additional mode into which light can be scattered. More specifically, to simulate the action of a balanced beamsplitter, assigning equal probabilities to the two outcomes for single-photon inputs, we consider the case with
$u_{11}=u_{21}=u_{12}=u_{22}=t\in\mathbb R$, and we use the notation $u_{31}= f_1, u_{32}= f_2$.
The request of overall unitarity imposes $\abs{f_1}=\sqrt{1-2t^2}$, and $f_2=-f_1$.
We can furthermore assume $f_1\in\mathbb R$, and thus finally obtain the following relations defining the lossy beamsplitter
\begin{equation}
\begin{aligned}
    a_1^\dagger \to t(a_1^\dagger + a_2^\dagger) + f_1 a_3^\dagger,\\
    a_2^\dagger \to t(a_1^\dagger + a_2^\dagger) - f_1 a_3^\dagger.
\end{aligned}
\end{equation}
Introducing the operator $a_+\equiv (a_1+a_2)/\sqrt2$, we can rewrite these as
\begin{equation}
    a_1^\dagger \to \sqrt2 t a_+^\dagger + f_1 a_3^\dagger, \quad
    a_2^\dagger \to \sqrt2 t a_+^\dagger - f_1 a_3^\dagger.
\end{equation}
A straightforward calculation then leads to
\begin{equation}
\begin{aligned}
    a_1^{\dagger 2} + (e^{i\phi} a_2^{\dagger})^{2} &\to
    2t^2(1+e^{i2\phi}) a_+^{\dagger 2} +
    f_1^2(1+e^{i2\phi}) a_3^{\dagger 2} + \\
    & + 2\sqrt2 t f_1(1-e^{i2\phi}) a_+^\dagger a_3^\dagger.
\end{aligned}
\end{equation}
Considering the unitary constraint at the maximal condition of the coherent absorption, we obtain $|f_1|=\sqrt2 t=1/\sqrt2$. In the case $\phi=q\pi+\pi/2,q \in \mathbb{Z}$ we have
\begin{equation}
    a_1^{\dagger 2} - a_2^{\dagger 2} \to
    2 a_+^\dagger a_3^\dagger.
\end{equation}
By introducing the states $\ket{n_j}$, which are Fock state of $n$ excitations in mode $j=1,2,3$, we have
\begin{equation}
    \frac{(\ket{2_1,0_2}-\ket{0_1,2_2})}{\sqrt2}\ket{0_3}\to
    \frac{\left(\ket{1_1,0_2}+\ket{0_1,1_2}\right)}{\sqrt 2}\ket{1_3},
\end{equation}
that is, one photon is deterministically absorbed while the other evolves into a balanced superposition of the two output modes. Such coherent absorption phenomenon can thus be thought of as an \textit{inverse} Hong-Ou-Mandel (HOM) effect between the input modes $a_1, a_2$ and the output modes $a_+,a_3$. In contrast, for $\phi=q\pi$, we have
\begin{equation}
    a_1^{\dagger 2} + a_2^{\dagger 2} \to
    a_+^{\dagger 2} + a_3^{\dagger 2},
\end{equation}
which corresponds to the state transformation
\begin{equation}
\begin{aligned}
        \frac{\ket{2_1,0_2}+\ket{0_1,2_2}}{\sqrt2}\ket{0_3}&\to \frac{(\ket{2_1,0_2}+\ket{0_1,2_2})}{2\sqrt 2}\ket{0_3} +\\ &+\frac{1}{2}\ket{1_1,1_2,0_3}+\frac{1}{\sqrt{2}}\ket{0_1,0_2,2_3}.
\end{aligned}
\end{equation}
This clearly shows that no single-photon absorption occurs in this case, while two-photon absorption takes place with a probability of $50\%$.

\begin{figure*}[htp]
\centering
\includegraphics[width=0.70\linewidth]{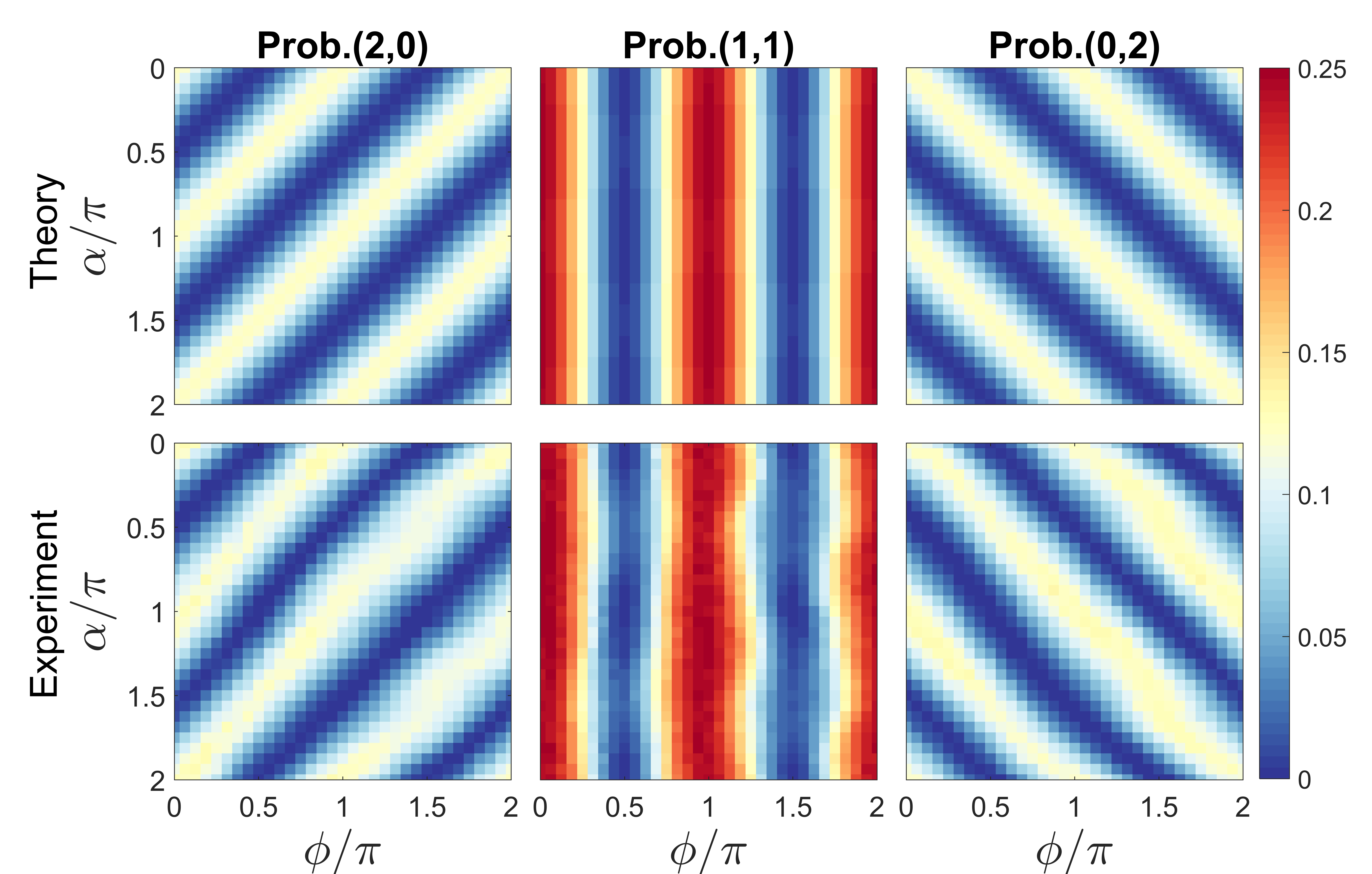}
\caption{Contributions of the two-photon survival probability: (Theory) top panel and (Experiment) bottom panel. Decomposition of the two-photon survival probability into its three contributions, $\text{Prob.}(2,0)$, $\text{Prob.}(0,2)$, and $\text{Prob.}(1,1)$, corresponding to two photons detected on the upper path, lower path, or on both paths, respectively. Each data point was integrated for 10 s. All probabilities are normalised with the probability of two-photon survival in a case of mimicking the lossless MZ interferometer with $\alpha=\pi/2$.}
\label{figS4}
\end{figure*}

The model presented above explains the maximum coherent absorption effect where the LTBS is programmed to $\alpha=0$. We highlight that the effect results from the non-unitarity of the LTBS. In the main text, we show full programmability of $\alpha$ -- from non-unitarity to mimicking unitarity -- resulting in a modulation of the two-photon survival probability. In Fig.~\ref{figS4}, we provide the three contributions of two-photon survival probability, $\mathrm{Prob.}(2,0)$, $\mathrm{Prob.}(0,2)$, $\mathrm{Prob.}(1,1)$. These correspond to the probabilities of two-photons being in mode 1, being in mode 2, or of one photon being in either mode, respectively. Experimentally, the two-photon survival probability is obtained by the sum of the probabilities of measured two-fold coincidences across all six output pairs, $2P_{\text{H1V1}}+P_{\text{H1H2}}+P_{\text{H1V2}}+P_{\text{V1H2}}+P_{\text{V1V2}}+2P_{\text{H2V2}}$. The multiplicative factor 2 at $P_{\text{H1V1}}$ and $P_{\text{H2V2}}$ takes the probability of two-photons bunching at the analysed beamsplitters into account. As shown in Fig.~\ref{figS4}, at $\alpha=p\pi, p \in \mathbb{Z}$ (the maximally lossy case), we observe in-phase oscillations of these contributions, which show the maximum two-photon survival probability of 0.5 when the two-photon N00N state is $1/\sqrt2(\ket{2_1,0_2}+\ket{0_1,2_2})$. In contrast, zero probability of two-photon survival is obtained when $\phi=q\pi+\pi/2,q \in \mathbb{Z}$. At $\alpha=\pi/2$ (mimicking the lossless case), the probabilities of having two photons in either path of the Mach-Zehnder (MZ) interferometer are out-of-phase to the probability of having exactly one photon on each path, resulting in a constant two-photon survival probability.

\newpage
\twocolumngrid

\end{document}